\documentclass[letterpaper]{article} 
\pdfoutput=1 
\usepackage{aaai2026}  
\usepackage{times}  
\usepackage{helvet}  
\usepackage{courier}  
\usepackage[hyphens]{url}  
\usepackage{graphicx} 
\usepackage{natbib}  
\usepackage{caption} 
\usepackage{booktabs}
\usepackage{amsmath} 
\graphicspath{{figures/}}
\nocopyright 

\title{Auditing Differential Visibility of Political Content on TikTok}
\author{
    Hazem Ibrahim\textsuperscript{\rm 1},
    Tewoflos Girmay\textsuperscript{\rm 1}
}
\affiliations{
    \textsuperscript{\rm 1}New York University Abu Dhabi, UAE\\
    hazem.ibrahim@nyu.edu
}

\begin{document}

\maketitle

\begin{abstract}
Allegations that TikTok shadow bans political content shape what creators post, what advertisers fund, and how regulators act, yet they are hard to adjudicate because platforms do not disclose how content is ranked. We test the claim with a dense hourly panel of $556{,}946$ follower-normalized view observations across $2{,}753$ videos from $67$ accounts curated into pro and anti sides of three contested topics (U.S.\ immigration enforcement, Trump coverage, and Israel/Palestine). On-topic videos are identified by a multi-step classifier, and stance is taken from each account's curated side. The conventional analysis appears to answer yes. Pooling the hourly snapshots, the topic-conditional reach gap reaches $p<10^{-140}$. Analyzed at the account level, the independent unit at which we sample accounts and assign stance, the gap disappears. Every account-level reach contrast is null after correction (BH-FDR $q\approx0.9$), with near-zero effect sizes. We find no evidence of moderate-to-large reach suppression on any topic. The null is informative. Account-level confidence intervals and a power analysis rule out such effects. As a design check, the same framework detects a clear asymmetry on a different outcome. Oppositional content (anti-Trump, pro-Palestine) earns more engagement per view rather than less reach (Cliff's $\delta=-0.51$ and $-0.64$; $q<0.03$). Higher engagement per view does not by itself rule out suppression, but it shows that the design can detect effects of this magnitude when they exist. The apparent reach gap is an artifact of two factors. The first is pseudoreplication, which counts tens of thousands of autocorrelated video-hours as independent observations; the second is confounding, since the side that looks suppressed is larger and, in the case of the Israel/Palestine topic, posts mostly in Arabic. In this corpus, what is taken for a shadow ban is better explained by a more engaged audience than by a suppressed one. We close with what a credible visibility audit requires.
\end{abstract}


\section{Introduction}
On creator forums the same accusation recurs, that the platform quietly throttles a user's reach because of what they post. On TikTok the charge attaches especially to political content, where users, journalists, and advocacy groups allege that one side of a contested issue is suppressed~\cite{savolainen2022shadow, cotter2023gaslighting,nicholas2022shedding}. These beliefs have concrete effects. They change what people post, drive advertiser and platform policy, and feed narratives of political censorship. Platforms do not disclose ranking, so the claim is argued from the outside by comparison, at every level of formality. Creators read a drop in their own view counts against their usual numbers as proof~\cite{nicholas2022shedding,delmonaco2024tiktok}, advocacy reports compare pooled hashtag totals across sides and platforms or compile user-reported takedowns~\cite{ncri2023timebomb,hrw2023meta}, and published audits formalize the same design, pooling posts from the accounts on each side and reporting the gaps that reach significance~\cite{yarchi2025imagewar}. What almost none of these comparisons fix is the unit of independent observation, and that choice, we show, decides the answer.

The measurement problem has consequences beyond research. Allegations of partisan suppression now feed congressional hearings, ad-spend decisions, and a fast-moving regulatory agenda that increasingly turns on independent audits of platform behavior~\cite{nicholas2022shedding, metaxa2021auditing}. When a headline is manufactured by an analysis choice rather than a platform action, it misdirects all three. Getting the method right is therefore part of the finding, and it is why we treat the gap between a significant-looking result and a defensible one as a result in its own right.

Here, we ask whether TikTok shadow bans political content, quietly suppressing one side of a contested issue because of the stance that content takes. To answer it, we tracked $2{,}753$ videos from $67$ curated political accounts across three contested topics, recording reach and engagement every hour, labeling on-topic videos with a human-validated ensemble LLM classifier, and taking each video's stance from its account's curated side. This process yielded $556{,}946$ hourly observations, a dense panel with a median of $99$ observations per video. We hypothesized that visibility might differ by stance, both because shadow-ban discourse predicts such a gap and because TikTok's recommender allegedly favored Republican-aligned content during the 2024 U.S.\ elections~\cite{ibrahim2026tiktok}. Running the literature's standard pooled analysis on our own data, the topic-conditional reach asymmetry appears at $p<10^{-140}$, comfortably surviving a multiple-testing correction. However, this detection does not survive scrutiny. Analyzed at the account level, the gap disappears. No topic shows a significant difference in reach, and the effect sizes are essentially zero. The design has the power to detect a reach gap. At the same sample size, the same test finds a clear difference in \emph{engagement}, with oppositional content earning more reactions per view, a direction inconsistent with a visibility penalty.

The two analyses disagree by more than a hundred orders of magnitude on identical data, and explaining the disagreement is the methodological contribution of the paper. The pooled $p<10^{-140}$ is manufactured by counting tens of thousands of autocorrelated hourly snapshots as independent observations, a textbook pseudoreplication, and the residual gap that remains is explained by account size and content language rather than stance. The pattern has a precedent in the account-level reanalysis that overturned the YouTube radicalization claim~\cite{hosseinmardi2021examining}.

Three contributions follow. First, we provide a dense, classifier-labeled TikTok audit of differential visibility on three contested political topics, and a clean account-level result of \textbf{no evidence of moderate-to-large reach suppression}, accompanied by confidence intervals and a power analysis that make the null informative and quantified (Sections~\ref{sec:did} and~\ref{sec:found}). Second, we identify the one robust signal in the data, an \textbf{engagement} asymmetry favoring oppositional content, which reframes perceived suppression in terms of audience intensity (Section~\ref{sec:found}). Third, we give a quantified account of \textbf{why the conventional analysis concludes otherwise}, tracing the spurious detection to pseudoreplication and stance-aligned confounds on the same corpus, and distill a short protocol for credible visibility audits (Section~\ref{sec:why}). We do not claim that platforms never differentially distribute content. We claim that on a high-quality panel, analyzed at the account level, the alleged effect is not present at the magnitudes the design can resolve, while a different and measurable asymmetry is present.

\section{Related work}\label{sec:related}
\subsection{Shadow banning, visibility moderation, and folk theories}
Shadow banning denotes a platform reducing a user's visibility without notice~\cite{lemerrer2021shadow}. Platforms increasingly moderate by demotion rather than removal, treating ``borderline'' content with reduced distribution~\cite{gillespie2022reduction,zuckerberg2018blueprint,goldman2021remedies}, which makes reduced reach a genuine governance lever but a hard one to observe from outside~\cite{zeng2022visibility,gillespie2018custodians,gorwa2020algorithmic}. Because ranking is opaque, users reason about it through folk theories assembled from personal observation~\cite{eslami2015algorithms,rader2015understanding}. The belief in being shadow banned is one such theory, an ``algorithmic folklore'' that platforms deny and users cannot readily disprove~\cite{savolainen2022shadow, cotter2023gaslighting}. On TikTok, the theory is bound up with identity. Creators infer that the platform buries content tied to marginalized identities or out-of-favor politics, and they adapt what they post to test and evade the suspected penalty~\cite{karizat2021algorithmic,simpson2021you,are2022shadowban,duffy2023margins}. The folk method of ``proof'' is comparative and widespread. In a representative U.S. survey, a drop in engagement metrics was the most common way users diagnosed their own shadowban ($44\%$), followed by checking visibility from a second account ($42\%$)~\cite{nicholas2022shedding}, and interview studies find users treating view-count gaps between comparable posts as definitive proof~\cite{delmonaco2024tiktok,cotter2023gaslighting}. That comparative logic is exactly what a pooled audit formalizes, and showing where it breaks is the purpose of this paper.

\subsection{Auditing algorithmic systems}
External audits are the standard tool for studying opaque platforms~\cite{sandvig2014auditing,metaxa2021auditing,bandy2021problematic, ibrahim2023YouTube, aldahoul2026schadenfreude}, ranging from sock-puppet personalization audits of search~\cite{hannak2013measuring, robertson2018auditing} to recommender audits of video platforms~\cite{hussein2020measuring}. A recurring lesson is that conclusions are fragile to how the data are analyzed, and that the unit of analysis often matters more than the size of the corpus. \citet{hosseinmardi2021examining} show that account-level consumption data overturn the popular claim that YouTube's algorithm drives radicalization. Account-level evidence has also reshaped the debate over political amplification. Analyzing a randomized holdout rather than an observational side-by-side, \citet{huszar2022amplification} estimate that Twitter's algorithm amplifies right-leaning content in six of seven countries studied, a real but bounded effect. The same study detects no significant association between an individual politician's party and amplification, so even with platform-side randomization the unit of analysis changes the answer. On TikTok, audits have characterized the For You feed's personalization factors~\cite{boeker2022empirical,vombatkere2024tiktok}, its political content~\cite{medinaserrano2020dancing,ibrahim2026tiktok}, and users' theories of how it boosts or buries posts~\cite{klug2021trick,bhandari2022algorithmized}.

Direct attempts to measure differential visibility vary widely in method, and the axis on which they vary most is the unit of analysis. At one end, high-profile advocacy comparisons use no independent unit at all. A widely circulated report benchmarks platform-lifetime post totals for sixty hashtags on TikTok against Instagram and reads the skewed ratios as evidence of likely deliberate suppression, with no uncertainty quantification and no adjustment for platform age, regional availability, or audience demographics~\cite{ncri2023timebomb}; its critics answered with placebo hashtags rather than statistics~\cite{matzko2024lies,harwell2023hashtags}. A descriptive analysis of Israel/Gaza content on TikTok reports side-level view totals whose mean and median disagree, the pro-Palestinian mean running nearly five times higher while its median is lower, the signature of a tail-driven pooled comparison~\cite{radway2025updating}. In the peer-reviewed literature the same comparative design appears at the post level. An audit of the May 2021 escalation compares $318$ hashtag-sampled TikTok videos across sides using independent-samples $t$-tests and pooled regressions, treating each video as an independent observation with no clustering by account, while the sides differ by an order of magnitude in mean follower count~\cite{yarchi2025imagewar}. At the other end sit account-level designs. \citet{lemerrer2021shadow} test shadow-ban detections account by account over $2.5$ million Twitter profiles and argue that much of the public evidence dissolves under careful definition and measurement; \citet{jaidka2023silenced} repeatedly test a stratified random sample of roughly $25{,}000$ U.S. accounts and find shadowbans rare and concentrated among bot-like accounts; recent sock-puppet audits aggregate exposure to the account before testing~\cite{ye2025auditing} or cluster errors by bot and week~\cite{ibrahim2026tiktok}. The literature, in short, straddles the unit-of-analysis line, and no external audit has to our knowledge shown on a single corpus that this choice alone manufactures or dissolves a detection. That demonstration is our contribution, on a dataset dense enough to remove the objection that the data are too thin.

The strongest evidence about platform effects comes from experiments run on the platform itself. In the 2020 U.S.\ Facebook and Instagram studies, researchers randomized feed ranking and composition for consenting users and measured downstream exposure, attitudes, and behavior~\cite{guess2023feed,nyhan2023likeminded, gonzalezbailon2023asymmetric}; switching to a chronological feed changed what users saw substantially but moved their attitudes little. These designs identify causal effects because treatment is assigned rather than inferred; absent platform cooperation, an external audit can characterize exposure and distribution but should report bounded, correlational conclusions.

\subsection{Measurement hazards}
Three known hazards shape any such audit. The first is \emph{pseudoreplication}, in which correlated repeated measurements are treated as independent replicates, inflating significance without adding information~\cite{hurlbert1984pseudoreplication,lazic2010pseudoreplication, lazic2018whatisn,galbraith2010clustered}; the remedies are to analyze at the independent unit or model the dependence, and to report the design effect and effective sample size~\cite{aarts2014solution,kish1965survey}. The second is \emph{researcher degrees of freedom}, the data-contingent analytic choices that manufacture significance even without intent~\cite{gelman2014statistical,simmons2011false}, a recognized driver of the broader replication crisis~\cite{ioannidis2005why, osc2015estimating}; the accepted defenses are pre-registration, full-grid reporting with multiplicity control, and specification-curve or multiverse analysis~\cite{simonsohn2020specification,steegen2016multiverse, benjamini1995controlling,nosek2018preregistration}. The third is \emph{measurement bias}. The language-identification and LLM stance classifiers now common in audits are less reliable on non-English, Global-South content~\cite{blasi2022systematic,joshi2020state,kreutzer2022quality}, and content classifiers carry social bias~\cite{sap2019risk,bender2021dangers}; LLM annotation is promising but non-deterministic and task-dependent, and must be validated against human labels~\cite{gilardi2023chatgpt,reiss2023testing,pangakis2023automated, ziems2024can}. Our design is built to address all three.

\section{Data and design}\label{sec:did}
TikTok distributes most views through the For You feed, a recommender that surfaces posts to users who do not follow the creator, so a post's reach is only loosely tied to its author's follower count~\cite{boeker2022empirical,vombatkere2024tiktok}. A shadow ban is therefore plausible in principle, the platform damping a post's distribution with no visible action, and the same opacity makes reach hard to measure from outside. We use follower-normalized plays as our reach proxy because it is the quantity the folk audits themselves use and the one a creator can observe, while noting that it confounds algorithmic distribution with audience behavior; only platform-side logs could separate the two, and we return to this limitation in the discussion.

\subsection{A dense hourly panel}
We tracked public TikTok accounts curated into pro and anti sides of three topics, U.S.\ immigration enforcement (ICE), Trump coverage (TRUMP), and Israel/Palestine. For each account we followed its recent posts, recording \texttt{playCount}, \texttt{diggCount} (likes), \texttt{commentCount}, \texttt{shareCount}, and \texttt{collectCount} (saves) every hour, alongside follower counts. Collection ran through TikAPI, a third-party wrapper over TikTok's public app endpoints. We polled each tracked post's public counters directly rather than sampling any personalized For You feed, so personalization does not enter the measurements. Over the collection window (3--28~April 2026) the collector completed $604$ hourly sweeps, and $598$ of the $602$ inter-sweep gaps are exactly one hour. The realized panel is therefore dense and uniform, with $556{,}946$ real snapshot rows over $2{,}753$ videos and $67$ accounts and a median of $99$ observations per video. Cumulative counters increment organically between consecutive hours, so each row is a fresh measurement rather than a carried-forward value; by the standards of prior audits, this is a high-quality panel, and the problems we document are not artifacts of thin data.

\subsection{Account selection and the analytic sample}\label{sec:sample}
The 67 accounts were curated before data collection began to ensure representation of both sides of each topic across a broad range of audience sizes. For each topic, we identified candidate accounts through Google searches for prominent voices on each side of the issue, classified them by their publicly expressed position, organized them into predefined follower-size tiers (from micro accounts to multi-million-follower creators), and selected accounts so that opposing sides contained comparable coverage across these strata. The resulting roster spans approximately 2.5K to 22M followers and includes the actors that shape political discourse on TikTok, from institutional and official accounts (party organizations, campaigns, militaries, news outlets) to commentators, independent creators, and on-the-ground journalists. Side membership reflects each account's publicly expressed position on the topic under study; accounts producing content on several issues appear in more than one topic, assigned by their position on each. The roster was finalized before hourly collection began, so selection is independent of the outcomes analyzed here. Table~\ref{tab:roster} lists these accounts in de-identified form; we withhold handles because naming public figures next to a stance and reach claim on these topics could be used to target them.

The follower-size tiers reduce structural differences in audience size rather than produce one-to-one matching, so the sides remain heterogeneous in follower count, posting cadence, language, and content; we measure and adjust for these differences analytically (Section~\ref{sec:why}) rather than eliminate them by design. The dataset is therefore a curated, stratified convenience sample, not a probability sample of political TikTok, which limits external validity (Appendix~\ref{sec:limits}). An account enters a topic-specific analysis only if it posted at least one relevant video during the window, so the accounts analyzed per topic (Table~\ref{tab:design}) number fewer than the curated roster. Appendix~\ref{app:accounting} reconciles every sample count reported in the paper, from the complete curated roster to each analytic subset.

\subsection{Stance and outcome measures}
Our reach measure is $\text{ViewsPer1K}=1000\cdot\texttt{playCount}/\text{followers}$, and engagement-per-view is total reactions divided by plays. Each video passed through a multi-step LLM classifier that first decides relevance, then topic, then a per-video stance ($\text{relevance}\rightarrow\text{topic}\rightarrow\text{stance}$). Of the $2{,}753$ tracked videos, $2{,}469$ were classifiable and $940$ are on-topic for one of the three issues; $754$ of these also carry a clear per-video stance (Table~\ref{tab:accounting}). Because LLM classifiers are contestable on non-English content~\cite{gilardi2023chatgpt,reiss2023testing,blasi2022systematic}, we deliberately limit the classifier's authority, using it only to identify which of an account's videos are on-topic and taking stance for the primary analysis from the account's curated side rather than from the per-video label. The account-level reach analysis therefore aggregates the $816$ on-topic videos belonging to a curated pro/anti account to one value per account (Table~\ref{tab:design}). Taking stance from the curated side insulates the headline result from the classifier's largest source of error, and the per-video labels agree with the curated side often enough to corroborate the curation (Section~\ref{sec:classval}). We report engagement-per-view pooled and decomposed into likes, comments, shares, and saves, since the components carry different signals about audience intensity.

\subsection{Classifier and its reliability}\label{sec:classval}
The classifier pipeline deployed is a three-model LLM ensemble (GPT-4o, Gemini, and Claude Sonnet) queried at temperature $0$, run relevance $\rightarrow$ topic $\rightarrow$ stance; the models vote, we keep the majority label, and the $26$ on-topic videos without a majority are excluded. The models read content in its original language, including Arabic, with no translation step. Inter-model agreement (Krippendorff's $\alpha$) is $0.81$ for topic and $0.78$ for stance, lower for relevance ($0.56$), which is why we use relevance only as a coarse filter. Because the headline estimand takes stance from the curated account side, classifier error enters the primary analysis only through which videos count as on-topic. As a check on the curation, the per-video stance label agrees with the curated side on $93.1\%$ of clearly-stanced videos, and the agreement does not degrade on Arabic ($93.4\%$ against $92.6\%$ on English), counter to the usual non-English reliability gap~\cite{blasi2022systematic,joshi2020state,kreutzer2022quality}; the per-topic and per-language breakdown is in Appendix~\ref{app:classval}. To validate the filter directly, we hand-labeled a stratified gold set of $200$ videos (balanced across topic, side, and language) plus $100$ videos on which the three models disagreed on relevance; the ensemble matches the human labels closely on both (on-topic inclusion $F_1=0.88$ stratified and $0.80$ on the hard disagreement cases; full metrics in Appendix~\ref{app:classval}). The low relevance $\alpha$ therefore reflects single-model disagreement that the majority vote resolves correctly rather than a biased on-topic filter. The same per-video labels drive the video-level rung of the pseudoreplication ladder (Section~\ref{sec:why}), where the reach null is unchanged.

\subsection{The account as the unit of analysis}
Because accounts are the units we sample and the level at which side labels are curated, our primary estimand is an account-level side difference in follower-normalized reach. We therefore aggregate videos to accounts and use account-level randomization inference, and we do not claim to identify within-account, post-level demotion. This makes the account the independent unit of inference; whether any platform treatment operates at the account or the post level is a separate question, which we revisit with a per-video, account-clustered model in Section~\ref{sec:why}. We analyze each topic separately, as the original claim of \emph{topic-conditional} asymmetry was framed (Table~\ref{tab:design}). ICE is near-empty as a content topic; we report it throughout but it cannot support inference, so the informative topics are Trump and Israel/Palestine.

\subsection{Estimators and pre-registration}
We summarize each video by its peak observed reach and collapse videos to the account by the median, giving one value per account. For each account-level set we report Cliff's $\delta$, a rank effect size in $[-1,1]$, with an account-bootstrap $95\%$ CI. The primary test is a Mann--Whitney $U$ on the account values; our preferred $p$-value comes from randomization inference, permuting stance labels across accounts $B=2\times10^4$ times, which needs no large-sample assumption. We additionally fit cluster-robust regressions with follower and age controls, while noting that cluster-robust standard errors are unreliable at our cluster counts~\cite{cameron2015practitioners,barr2013keepitmaximal}. The analysis grid (3~topics~$\times$~3~outcomes), the account-level unit, the estimators, and the discovery/confirmatory split were fixed in a pre-registration frozen before any inferential test was run; deviations are logged with it. Following that plan we correct each outcome family with Benjamini--Hochberg~\cite{benjamini1995controlling, nosek2018preregistration}, because the three outcomes answer three distinct questions (reach, engagement, velocity) rather than one; we report a stricter global correction across the full grid as a sensitivity in Section~\ref{sec:found}, and the conclusions do not change.

To calibrate the procedures themselves, rather than only the data, we run a constructed-null simulation. We repeatedly reassign the stance labels at random across accounts ($3{,}000$ draws), which destroys any true stance effect by construction, and record how often each test rejects at $\alpha=0.05$. A correctly specified test should reject about $5\%$ of the time; a test that rejects far more often is detecting the data's dependence structure, not a stance effect. We run this calibration at each unit of analysis, which lets us measure directly how pooling inflates the false-positive rate (Section~\ref{sec:why}).

\section{An engagement gap, but no reach suppression}\label{sec:found}
\begin{figure*}[t]
\centering
\includegraphics[width=0.95\textwidth]{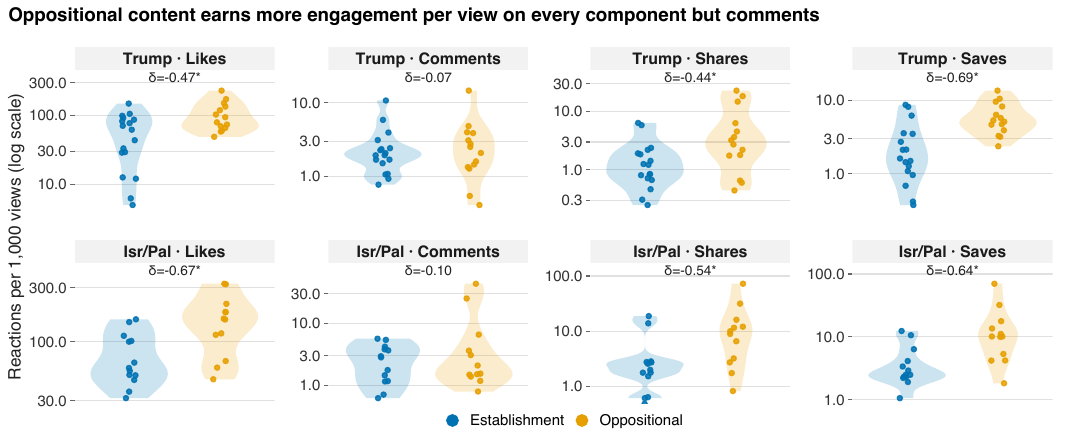}
\caption{Oppositional content earns more reactions per view, a direction inconsistent
with a visibility penalty. Per-account rates by side (log scale), annotated with the
account-level Cliff's $\delta$ ($\ast$ marks $p<0.05$); the gap appears in likes,
shares, and saves while comments differ little. Establishment denotes pro-Trump and
pro-Israel accounts, oppositional their opposites.}
\label{fig:engagement}
\end{figure*}
\subsection{A gap that vanishes at the account level}
The discourse and the standard analysis alike predict a topic-condi\-tional reach gap, and at first our data appear to deliver one. Pooling the hourly snapshots and comparing sides with a rank test, the Israel/Palestine reach gap reaches $p=6.7\times10^{-147}$ and the Trump gap falls below floating-point precision. The significance is already large at the video level (Trump $\delta=0.40$, $p=4\times10^{-10}$; Israel/Palestine $\delta=0.24$, $p=4\times10^{-4}$), and grows by scores of orders of magnitude when the hourly snapshots are added. By the conventions of the pooled designs reviewed in Section~\ref{sec:related}, large $N$, a vanishing $p$, and a passed multiple-testing correction, this meets the standard for a detection~\cite{yarchi2025imagewar}.

This apparent detection does not survive the move to account-level analysis. At the account level no reach comparison is significant (ICE $\delta=0.07$, $p=0.93$; TRUMP $\delta=0.08$, $p=0.71$; Israel/Palestine $\delta=0.06$, $p=0.81$), with BH-corrected $q\approx0.93$ throughout and randomization inference in agreement (Table~\ref{tab:grid}). The point estimates are near zero throughout ($|\delta|\le0.08$).

\subsection{The null is informative}
A null is only as strong as the design's power to detect an effect. The same design and sample size detect a clear effect on a different outcome. Engagement-per-view is lower for the establishment side on both informative topics, surviving the pre-registered correction (TRUMP $\delta=-0.51$, $q=0.025$; Israel/Palestine $\delta=-0.64$, $q=0.021$), with bootstrap CIs that exclude zero ($[-0.82,-0.15]$ and $[-0.94,-0.26]$). We use \emph{establishment} as descriptive shorthand for the pro-Trump and pro-Israel side and \emph{oppositional} for the anti-Trump and pro-Palestine side, with no normative judgment intended. We treat the engagement effect as a secondary, supporting result rather than a headline. It also survives a correction across the six informative-topic cells ($q=0.049$, $0.043$); only under the strictest possible correction, a global BH across all nine grid cells, does it turn borderline, with $q$ rising to $0.073$ (Trump) and $0.064$ (Israel/Palestine). Its role is to show that the design can resolve an effect of this size at this sample, which is what makes the reach null informative.

Two further analyses make the bound explicit. First, a Monte-Carlo power analysis shows that at the observed account counts the reach test has $76$--$89\%$ power to detect a true effect of $\delta=0.6$ and $63$--$70\%$ at $\delta=0.5$, the range of the engagement effect we do detect. Second, the account-bootstrap intervals for reach turn the null into a quantified equivalence statement. We can rule out a reach effect larger than $|\delta|=0.49$ on Trump and $|\delta|=0.51$ on Israel/Palestine at $95\%$ confidence (Table~\ref{tab:grid}), while smaller effects remain possible; the null is therefore bounded and quantified. The reach nulls also reproduce across pre-specified discovery (3--15~Apr) and confirmatory (16--28~Apr) windows, with none of the strengthening under confirmation that a real effect would produce (Appendix~\ref{app:robust}).

\begin{table}[t]
\centering
\caption{No reach effect survives at the account level, while the engagement effect
survives on both informative topics with a CI that excludes zero. Pre-registered grid
with account-bootstrap $95\%$ CIs, rank-test $p$, BH-corrected $q$ (within outcome
family), and randomization $p$.}
\label{tab:grid}
\setlength{\tabcolsep}{4pt}
\resizebox{\columnwidth}{!}{%
\begin{tabular}{llccccc}
\toprule
Topic & Outcome & $n_a/n_b$ & $\delta$ [95\% CI] & $p_{\text{MW}}$ & $q_{\text{BH}}$ & $p_{\text{perm}}$ \\
\midrule
ICE     & reach    & 6/5   & $+0.07$ $[\text{-}0.67,0.73]$ & $0.93$  & $0.93$ & $0.86$ \\
TRUMP   & reach    & 17/14 & $+0.08$ $[\text{-}0.34,0.49]$ & $0.71$  & $0.93$ & $0.75$ \\
ISR/PAL & reach    & 13/12 & $+0.06$ $[\text{-}0.41,0.51]$ & $0.81$  & $0.93$ & $0.91$ \\
ICE     & eng/view & 6/5   & $-0.13$ $[\text{-}0.80,0.60]$ & $0.79$  & $0.79$ & $1.00$ \\
TRUMP   & eng/view & 17/14 & $-0.51$ $[\text{-}0.82,\text{-}0.15]$ & $0.016$ & $0.025$& $0.015$ \\
ISR/PAL & eng/view & 13/12 & $-0.64$ $[\text{-}0.94,\text{-}0.26]$ & $0.007$ & $0.021$& $0.002$ \\
ICE     & velocity & 6/5   & $+0.13$ $[\text{-}0.67,0.80]$ & $0.79$  & $0.79$ & $0.81$ \\
TRUMP   & velocity & 17/14 & $-0.23$ $[\text{-}0.62,0.19]$ & $0.29$  & $0.44$ & $0.21$ \\
ISR/PAL & velocity & 13/12 & $+0.42$ $[\text{-}0.01,0.80]$ & $0.077$ & $0.23$ & $0.26$ \\
\bottomrule
\end{tabular}}
\end{table}

\subsection{An engagement asymmetry, not reduced reach}
The engagement asymmetry is the clearest signal in the data, and it points the opposite way from a visibility penalty. Oppositional content earns more per-view likes, shares, and saves on both informative topics, while comments, themselves an effortful reaction, differ little (Figure~\ref{fig:engagement}). The per-component gaps are consistent and sizable. On Trump, $\delta=-0.47$ for likes, $-0.44$ for shares, and $-0.69$ for saves, against $-0.07$ for comments; on Israel/Palestine, $-0.67$, $-0.54$, and $-0.64$ against $-0.10$ for comments. Saves, among the most deliberate signals a viewer can send, show the largest gap on both topics. Higher engagement per view does not by itself rule out reach suppression, since a platform could in principle demote content that draws strong reactions; but paired with the reach null, the pattern is more consistent with audience mobilization than with reach suppression, in line with the long-noted gap between counted engagement and circulation~\cite{gerlitz2013like,cotter2019visibility}. A pooled video-level comparison of the May 2021 escalation likewise reported higher engagement on pro-Palestinian posts~\cite{yarchi2025imagewar}; at the account level, that engagement asymmetry survives the unit correction while the reach suppression the discourse infers from such gaps does not. The engagement gap also persists under an English-only restriction, whereas the apparent reach gap, as we show next, does not.

\section{Why the conventional analysis finds otherwise}\label{sec:why}
The same data yield ``$p<10^{-140}$'' and ``null'' depending on a single choice, the unit of analysis. Two mechanisms account for the gap.

\subsection{Pseudoreplication}
Repeated hourly snapshots of one video are near-duplicates. Decomposing the variance of $\log(\text{ViewsPer1K})$, snapshots within a video have an intraclass correlation of $0.98$ (Appendix~\ref{app:variance}). Therefore, the $173{,}050$ on-topic snapshot rows carry the information of about $960$ independent videos (design effect~$\approx180$)~\cite{kish1965survey,killip2004icc}, and those videos nest within only the few dozen accounts where stance varies. Counting the snapshots as independent observations, as a pooled test does, treats an effective sample of dozens as one of hundreds of thousands~\cite{hurlbert1984pseudoreplication,lazic2018whatisn, aarts2014solution}. The per-topic effective sample is $471$ videos for Trump and $460$ for Israel/Palestine (Appendix~\ref{app:variance}), not the tens of thousands of rows, and those videos in turn nest within $31$ and $25$ accounts. Because a $p$-value shrinks with $\sqrt{N}$ while a rank effect size does not, inflating the unit drives the $p$-value toward zero while leaving $\delta$ essentially where it was.

\begin{figure*}[t]
\centering
\includegraphics[width=0.95\textwidth]{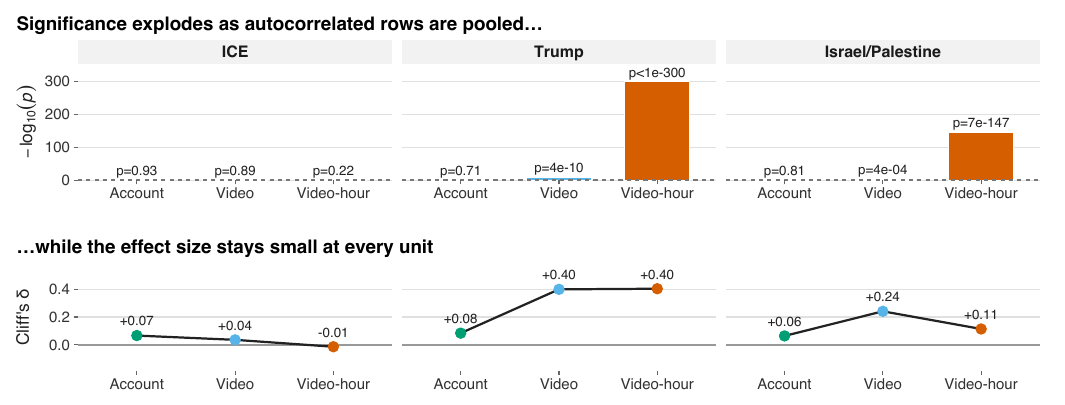}
\caption{Inflating the unit of analysis manufactures significance. The same reach
comparison at three units; the $p$-value (top) grows without bound as autocorrelated
snapshots are counted as independent, while Cliff's $\delta$ (bottom) stays small at
every unit.}
\label{fig:artifact}
\end{figure*}

Figure~\ref{fig:artifact} shows the consequence. For each topic we run the same reach comparison at the account, video, and video-hour units. As the unit inflates, the $p$-value collapses while the effect size remains essentially unchanged. For Israel/Palestine the account-level gap is null ($\delta=0.06$, $p=0.81$) yet reaches $p=6.7\times10^{-147}$ at the video-hour level, with $\delta$ never exceeding $0.24$. A constructed-null simulation makes the error explicit. When we randomly reassign stance to accounts, destroying any true effect, the account-level test holds near its nominal rate ($4\%$ pooled, $3$--$6\%$ by topic), while the pooled video-hour test rejects $95\%$ of the time (Figure~\ref{fig:fprunit}; per-topic detail in Appendix~\ref{app:robust}). Under a constructed null with no stance effect at all, the pooled procedure still reports significance almost every time, so the reported detection is manufactured outright. The pooled design is not a strawman. A peer-reviewed TikTok comparison of this same conflict treats hashtag-sampled videos as independent observations in exactly this way~\cite{yarchi2025imagewar}, and the comparisons that reached congressional testimony pool hashtag aggregates with no unit at all~\cite{ncri2023timebomb}.

\subsection{Confounds aligned with stance}
\begin{figure*}[t]
\centering
\includegraphics[width=0.95\textwidth]{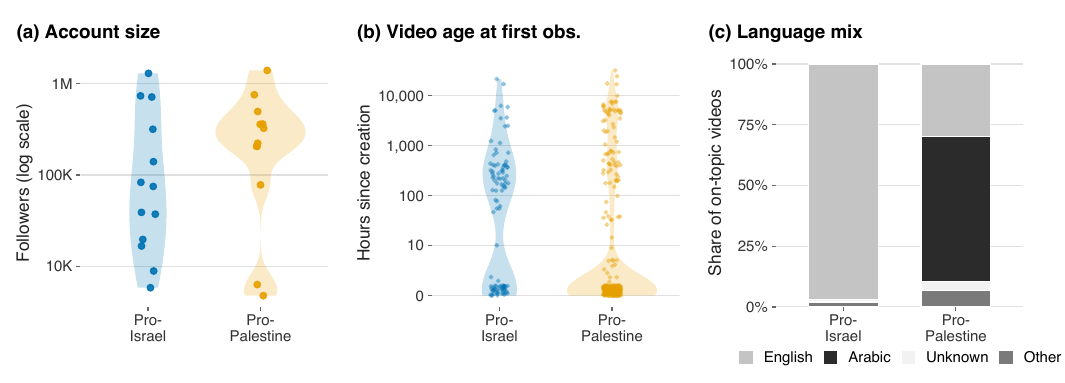}
\caption{Stance is confounded with (a) account size, (b) video age at first
observation, and (c) content language on the Israel/Palestine topic, so a
between-stance reach gap pooled below the account level is not identifiable.}
\label{fig:confounds}
\end{figure*}

The residual gap that pooling exaggerates is not stance to begin with, because the sides are not comparable on three nuisance dimensions. The first is \emph{size}. Reach is sublinear in followers (the slope of $\log\texttt{playCount}$ on $\log\text{followers}$ is $0.73$ for Trump and near zero for Israel/Palestine, both below one), so the per-follower metric over-penalizes large accounts, and the pro-Palestine accounts are $3.6\times$ larger than pro-Israel accounts at the median (Figure~\ref{fig:confounds}). The second is \emph{age}. Pro-Israel videos enter our panel far later in their life cycle (median $128$ hours since creation, against under one hour for pro-Palestine), so the two sides are compared at different points on the view-accumulation curve. The third is \emph{language}. Pro-Palestine accounts post mostly Arabic content and pro-Israel accounts almost none, while collection defaulted to a U.S.\ region, so stance is nearly collinear with content language and audience geography~\cite{blasi2022systematic,sap2019risk,kreutzer2022quality}. The collector ran with a default U.S.\ country setting; because the public counters are global totals, this does not change a video's numbers, only which videos are discoverable and the language environment in which a U.S.\ viewer would meet them, so the confound is one of content and audience composition rather than of measurement. Restricting to English-only videos, the account-level reach contrast stays null ($\delta=-0.21$, $p=0.42$; Figure~\ref{fig:lang}). There is no account-level reach gap for language to explain in the first place; the pooled gap that does appear is partly a comparison of Arabic content delivered to a U.S.\ audience against English content. The engagement gap, by contrast, persists within English and in fact strengthens slightly on the restricted set (engagement-per-view $\delta=-0.76$ for English-only, against $-0.64$ for all videos; Figure~\ref{fig:lang}), so it reflects how the two audiences react rather than a language artifact.

\subsection{Adjusting the confounds directly}
Adjusting for a confound in a model agrees with subsetting it away. We fit a per-video model of $\log(\text{peak plays})$ on side, with $\log(\text{followers})$ and $\log(\text{video age})$ as controls and standard errors clustered by account. The post is the observation but the account remains the unit of inference, the specification appropriate if moderation acts at the post level. After these controls no topic shows a significant side difference in reach (Trump $\beta=0.84$, $p=0.07$; Israel/Palestine $\beta=0.45$, $p=0.46$; ICE $\beta=0.34$, $p=0.70$); the single borderline cell, Trump, points if anything toward the establishment side rather than against the oppositional one. Because cluster-robust errors are anti-conservative at these cluster counts~\cite{cameron2015practitioners}, we recompute each side coefficient with a wild-cluster bootstrap (Rademacher weights, null imposed, $B=9{,}999$; Appendix~\ref{app:postlevel}). The correction removes even the Trump signal, whose bootstrap $p$ rises from $0.07$ to $0.22$, and no topic is significant under the small-cluster-aware test. The post-level adjusted analysis thus reaches the same place as the account-level test, with no robust reach gap.

Two targeted subset checks, restricting to videos first observed within $24$h ($48$h) of posting and stratifying the English-only topics by language, leave the reach nulls unchanged (Appendix~\ref{app:robust}).

\subsection{The result is fragile across the specification space}
\begin{figure}[t]
\centering
\includegraphics[width=0.96\columnwidth]{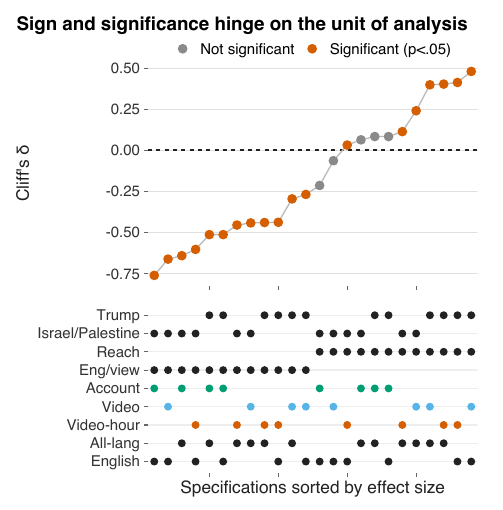}
\caption{Sign and significance of the reach ``effect'' flip across the $24$
defensible specifications, driven mainly by the unit of analysis. Each point (top) is
one analysis; its choices are shown below.}
\label{fig:speccurve}
\end{figure}

The choices that flip the result, the unit of analysis, the outcome metric, and the language subset, are each individually defensible, and an analyst rarely commits to all of them in advance~\cite{gelman2014statistical,simmons2011false}. Across $24$ defensible specifications (Figure~\ref{fig:speccurve}), $19$ are ``significant'' and the sign is mixed, but the variation is structured; significance is governed almost entirely by the unit of analysis, rising from $50\%$ of account-level specifications to $100\%$ at the video-hour level~\cite{simonsohn2020specification, steegen2016multiverse,silberzahn2018many}. Fix the unit at the account and the ambiguity collapses, with the reach question consistently null and the engagement question consistently positive.

\subsection{Toward credible visibility audits}
From these mechanisms we distill a minimal practice. (1)~Analyze real observations at the level of the independent unit, and report the design effect and effective $N$ before any pooled test. (2)~Make the sampled independent unit explicit; for account-curated side comparisons like ours it is the account, and if the hypothesized treatment is post-level, use hierarchical or cluster-aware models rather than treating repeated posts or snapshots as independent. (3)~De-confound account size, video age, and content language and region before attributing any gap to stance. (4)~Pre-register the grid and publish a specification curve rather than a single favorable cell. (5)~Report power and frame the result as a bounded claim, reserving causal claims for randomized on-platform experiments.

\section{Discussion}
Anyone who pools the snapshots can produce a significant-looking shadow-ban result from these data, complete with a vanishingly small $p$. Producing a defensible result is slower and, here, ends in a reach null. That asymmetry of effort has consequences beyond research, because journalists and advocacy organizations run effectively the same observational comparisons~\cite{ncri2023timebomb,radway2025updating,harwell2023hashtags}, and the incentives favor the significant-looking result. Our constructed-null and specification analyses quantify how easily that result can be manufactured, and in which direction. The mechanism involves no fraud, only the ordinary accumulation of defensible-looking choices. An analyst who collects hourly snapshots because dense data seem better, pools them because more rows seem better, and reports the comparison that reaches significance never once does anything that looks like p-hacking, yet arrives at a result inflated by two orders of magnitude.

The one asymmetry we do find, more sharing and saving of oppositional content, is consistent with audience mobilization rather than suppression. Saves and shares, the most effortful reactions a viewer can give, show the largest gaps; comments barely differ. The effect is correlational and may reflect who follows these accounts rather than how the platform treats them, and we do not claim otherwise. The methodological lesson is more general. Reach is a single, noisy measure of a post's reception, and an audit that measures the full interaction profile can find a real signal exactly where the reach comparison finds a spurious one.

None of this shows that platforms are even-handed; differential distribution may exist and matter greatly. The point is that establishing it requires either randomized on-platform experiments or observational designs with matched accounts, dense and uniform sampling, language and region held fixed, and analysis at the account level with pre-registration. Account-level care has overturned algorithmic-amplification claims before~\cite{hosseinmardi2021examining}; on these three topics, applied to TikTok reach, it does so again.

\section{Limitations}\label{sec:limits}
Several limitations bound our claims. Our corpus is observational, single-platform, and built from curated accounts, so we characterize the accounts we tracked, not TikTok as a whole. The account counts per topic are in the tens, so we bound rather than exclude reach effects smaller than the design resolves (Section~\ref{sec:found}). Reach is measured by follower-normalized plays, a proxy for distribution rather than a direct read of the ranker, and it cannot separate algorithmic demotion from audience behavior; only platform-side logs could. Our language analysis relies on imperfect platform language tags, which only sharpens the difficulty of separating stance from language. Stance is assigned at the account level from a curated side, our sampling unit, though it misses within-account variation across posts. The on-topic filter is an LLM ensemble; because the headline stance comes from the curated side, classifier error enters only through which videos count as on-topic, a gate we validate against a human-labeled gold set ($F_1\approx0.80$ even on the model-disagreement cases; Section~\ref{sec:classval}). That gold set is author-labeled, which bounds the check. Finally, ICE is too sparse to support inference and is reported for completeness only. None of these limitations is repaired by adding more autocorrelated rows, which is the central point of the paper.

\section{Ethics and data}
This study uses only data that TikTok exposes publicly, the counts of plays, likes, comments, shares, and saves for posts from public accounts, together with public follower counts. We did not interact with users or alter their feeds. We report results only in aggregate, at the level of curated sides and topics; released artifacts give per-account quantities only in de-identified, hashed form and never join a named handle to a stance attribution or a suppression claim. The topics, especially Israel/Palestine, are sensitive, and a per-account stance-and-reach table could be repurposed to target individuals. Stance labels describe an account's public, curated position, not any inference about a person's identity. Because measurement instruments such as language identification and LLM stance classifiers are known to be biased against non-English and Global-South content~\cite{blasi2022systematic,sap2019risk}, we keep those instruments out of the headline estimand and validate the ones we do use against human labels. Because this study analyzes only public, aggregate platform counters and involves no interaction with users, it did not require IRB review under our institution's criteria for research on public, non-interactive data.

\section{Conclusion}
We asked whether TikTok shadow bans political content, using a dense hourly panel built to answer the question. Analyzed conventionally, the data ``detect'' suppression with overwhelming significance. Analyzed at the unit where stance is defined, and after accounting for size and language confounds, they show no evidence of moderate-to-large differential reach suppression across three contested topics. The one robust asymmetry is greater audience engagement for oppositional content, with reach unchanged. Detecting differential visibility from public data is harder than current practice assumes, and in this corpus what is read as a shadow ban is better explained by a more engaged audience than by a suppressed one.

\bibliography{references} 

\appendix

\section{Sample accounting}\label{app:accounting}
Table~\ref{tab:design} gives the per-topic account-level design, and Table~\ref{tab:accounting} reconciles every sample count used in the paper, from the full tracked panel to each analytic subset. Table~\ref{tab:roster} lists the analytic accounts individually in de-identified form.

\begin{table}[h]
\centering
\caption{Per-topic account-level design. Reach observations are real snapshots; the
unit of analysis is the account.}
\label{tab:design}
\setlength{\tabcolsep}{4pt}
\resizebox{\columnwidth}{!}{%
\begin{tabular}{llccc}
\toprule
Topic & Side & Accounts & Videos & Median followers \\
\midrule
ICE    & \textsc{pro}          &  6 &   8 & 82K \\
ICE    & \textsc{against}      &  5 &  13 & 274K \\
TRUMP  & \textsc{pro}          & 17 & 143 & 224K \\
TRUMP  & \textsc{against}      & 14 & 246 & 326K \\
ISR/PAL& \textsc{pro-israel}   & 13 & 101 & 75K \\
ISR/PAL& \textsc{pro-palestine}& 12 & 305 & 274K \\
\bottomrule
\end{tabular}}
\end{table}

\begin{table}[h]
\centering
\caption{Sample accounting, from the full panel to each analytic subset. Counts that
appear elsewhere in the paper are reconciled here; ``--'' marks a count we do not use.
The account-level reach set attributes each on-topic video to its account's curated side
(Section~\ref{sec:sample}); the clearly-stanced set uses the per-video classifier label.}
\label{tab:accounting}
\small
\setlength{\tabcolsep}{4pt}
\resizebox{\columnwidth}{!}{%
\begin{tabular}{lccc}
\toprule
Stage & Videos & Snap.\ rows & Handles \\
\midrule
All tracked public videos        & $2{,}753$ & $556{,}946$ & $65^{\ast}$ \\
Classifiable by the ensemble     & $2{,}469$ & --          & -- \\
On-topic (one of three issues)   & $940$     & $173{,}050$ & $59$ \\
Strict inclusion (non-ambiguous) & $900$     & --          & -- \\
Clearly stanced (per-video)      & $754$     & --          & -- \\
Account-level reach set          & $816$     & --          & $57^{\dagger}$ \\
\bottomrule
\end{tabular}}
\\[2pt]
{\footnotesize The \emph{Handles} column counts distinct posting handles, so each row is
bounded by the row above it. The curated roster is $67$ accounts ($94$
handle$\times$topic cells); $^{\ast}$$65$ of them posted $\ge1$ tracked video.
$^{\dagger}$the reach set's $57$ handles form the $67$ account$\times$topic cells
analyzed per topic (Table~\ref{tab:design}).}
\end{table}

\begin{table*}[t]
\centering
\caption{De-identified roster of the $57$ accounts underlying the account-level tests,
one row per account. Handles are withheld and follower counts are banded so that no
account can be joined to a stance and a reach claim; per-topic-side membership counts
reproduce Table~\ref{tab:design}.}
\label{tab:roster}
\footnotesize
\setlength{\tabcolsep}{5pt}
\begin{tabular}[t]{cccl}
\toprule
ID & Tier & Followers & Side(s) \\
\midrule
A01 & 3 & 100K--1M & IP$\cdot$p \\
A02 & 3 & 100K--1M & TR$\cdot$a \\
A03 & 3 & 100K--1M & IP$\cdot$p \\
A04 & 1 & 10--100K & IP$\cdot$i \\
A05 & 1 & 10--100K & IP$\cdot$i \\
A06 & 3 & 100K--1M & IP$\cdot$p \\
A07 & 1 & $<$10K & TR$\cdot$p \\
A08 & 1 & $<$10K & IP$\cdot$p \\
A09 & 6 & 1--10M & TR$\cdot$p \\
A10 & 3 & 100K--1M & TR$\cdot$a \\
A11 & 1 & 10--100K & TR$\cdot$a \\
A12 & 3 & 100K--1M & IP$\cdot$p \\
A13 & 5 & 1--10M & TR$\cdot$a \\
A14 & 2 & 100K--1M & TR$\cdot$p \\
A15 & 3 & 100K--1M & TR$\cdot$p \\
A16 & 2 & 10--100K & TR$\cdot$a \\
A17 & 2 & 10--100K & IP$\cdot$p \\
A18 & 4 & 100K--1M & IP$\cdot$p \\
A19 & 1 & $<$10K & IP$\cdot$p \\
A20 & 2 & 10--100K & IP$\cdot$i \\
A21 & 4 & 100K--1M & IP$\cdot$i \\
A22 & 4 & 1--10M & TR$\cdot$a \\
A23 & 3 & 100K--1M & IP$\cdot$i \\
A24 & 5 & 1--10M & TR$\cdot$a \\
A25 & 1 & $<$10K & TR$\cdot$p \\
A26 & 3 & 100K--1M & IC$\cdot$a, TR$\cdot$a \\
A27 & 4 & 1--10M & IP$\cdot$i \\
A28 & 1 & $<$10K & IP$\cdot$i \\
A29 & 1 & 10--100K & IC$\cdot$p \\
\bottomrule
\end{tabular}
\hspace{2.5em}
\begin{tabular}[t]{cccl}
\toprule
ID & Tier & Followers & Side(s) \\
\midrule
A30 & 3 & 100K--1M & TR$\cdot$p \\
A31 & 1 & 10--100K & TR$\cdot$a \\
A32 & 2 & 10--100K & IP$\cdot$i \\
A33 & 2 & 10--100K & IC$\cdot$p, IP$\cdot$i \\
A34 & 2 & 10--100K & IP$\cdot$i \\
A35 & 4 & 1--10M & TR$\cdot$p \\
A36 & 5 & 1--10M & IC$\cdot$p, TR$\cdot$p \\
A37 & 6 & $>$10M & TR$\cdot$p \\
A38 & 3 & 100K--1M & IP$\cdot$i \\
A39 & 1 & 10--100K & TR$\cdot$a \\
A40 & 1 & $<$10K & IC$\cdot$p, TR$\cdot$p \\
A41 & 2 & 10--100K & IC$\cdot$a, TR$\cdot$a \\
A42 & 5 & 1--10M & IC$\cdot$a, TR$\cdot$a \\
A43 & 2 & 10--100K & TR$\cdot$p \\
A44 & 2 & 10--100K & TR$\cdot$p \\
A45 & 3 & 100K--1M & IP$\cdot$p \\
A46 & 3 & 100K--1M & IP$\cdot$p \\
A47 & 1 & 10--100K & IC$\cdot$a \\
A48 & 6 & $>$10M & TR$\cdot$p \\
A49 & 4 & 100K--1M & IC$\cdot$p, TR$\cdot$p \\
A50 & 3 & 100K--1M & TR$\cdot$p \\
A51 & 2 & 10--100K & IC$\cdot$p, TR$\cdot$p \\
A52 & 3 & 100K--1M & IC$\cdot$a, TR$\cdot$a \\
A53 & 4 & 1--10M & IP$\cdot$p \\
A54 & 3 & 100K--1M & IP$\cdot$p \\
A55 & 4 & 100K--1M & TR$\cdot$p, IP$\cdot$i \\
A56 & 1 & $<$10K & IP$\cdot$i \\
A57 & 2 & 10--100K & TR$\cdot$a \\
\bottomrule
\end{tabular}
\\[4pt]
{\footnotesize Topic codes IC (ICE), TR (Trump), IP (Israel/Palestine); side $\cdot$p
denotes the pro side (pro-Palestine under IP), $\cdot$a the anti side, and $\cdot$i
pro-Israel. Tier runs from $1$ (smallest) to $6$ (largest) follower stratum. Each
account appears once and lists every topic it was curated into, so per-topic-side
membership counts reproduce Table~\ref{tab:design}.}
\end{table*}

\section{Variance components and effective sample size}\label{app:variance}
Table~\ref{tab:icc} decomposes the variance of $\log(\text{ViewsPer1K})$ within videos on the real on-topic snapshots. Repeated hourly snapshots are near-duplicates, so the per-video effective sample size is roughly two orders of magnitude below the row count, and these videos nest within the few dozen accounts of Table~\ref{tab:design}.

\begin{table}[h]
\centering
\caption{Variance components for $\log(\text{ViewsPer1K})$ within videos on the real
on-topic snapshots, reporting the intraclass correlation (ICC), design effect (Deff),
and effective sample size ($N_\text{eff}$).}
\label{tab:icc}
\resizebox{\columnwidth}{!}{%
\begin{tabular}{lccccc}
\toprule
Topic (within video) & Videos & Rows & ICC & Deff & $N_\text{eff}$ \\
\midrule
All on-topic     & 940 & 173{,}050 & 0.98 & 180 & 963 \\
ICE              &  32 &  11{,}159 & 0.97 & 339 &  33 \\
TRUMP            & 458 &  71{,}717 & 0.97 & 152 & 471 \\
Israel/Palestine & 450 &  90{,}174 & 0.98 & 196 & 460 \\
\bottomrule
\end{tabular}}
\end{table}

\section{Classifier reliability detail}\label{app:classval}
The per-video stance label agrees with the curated account side on $93.1\%$ of the clearly-stanced videos for which both are defined ($637$ of $684$). The agreement is stable across topics (ICE $94.1\%$, Israel/Palestine $95.8\%$, Trump $90.3\%$) and across languages (English $92.6\%$ of $514$ videos, Arabic $93.4\%$ of $137$), so the per-video labels do not degrade on non-English content. The residual disagreement is dominated by the expected case of an account occasionally posting off its usual side rather than by any systematic, language-aligned error. Inter-model reliability across the three ensemble members, by Krippendorff's $\alpha$ on nominal labels, is $0.56$ for relevance, $0.81$ for topic, and $0.78$ for stance.

\paragraph{Human validation.}
To validate the classifier against human judgment we hand-labeled a gold set comprising a stratified random sample of $200$ videos (balanced across topic, side, and language) and an additional $100$ videos drawn from the cases where the three models disagreed on relevance, the hardest inputs for the ensemble. Table~\ref{tab:humanval} scores the ensemble majority vote against these labels. Agreement is high on the stratified set and stays strong on the disagreement cases; the most policy-relevant number, the on-topic inclusion decision that builds the analytic sample, matches the human labels at $F_1=0.80$ (disagreement).

\begin{table}[h]
\centering
\caption{Human validation of the ensemble classifier against an author-labeled gold set
of $200$ stratified videos plus $100$ videos where the models disagreed on
relevance. $F_1$ is for the positive/on-topic class (binary tasks) or macro-averaged
(multiclass, marked $^{m}$); $\kappa$ is Cohen's $\kappa$ (binary tasks).}
\label{tab:humanval}
\small
\setlength{\tabcolsep}{4pt}
\resizebox{\columnwidth}{!}{%
\begin{tabular}{lccc|ccc}
\toprule
 & \multicolumn{3}{c}{Stratified $200$} & \multicolumn{3}{c}{Disagreement $100$} \\
\cmidrule(lr){2-4}\cmidrule(lr){5-7}
Task & Acc & $F_1$ & $\kappa$ & Acc & $F_1$ & $\kappa$ \\
\midrule
Relevance (\textsc{yes/no}) & $0.91$ & $0.91$ & $0.88$ & $0.82$ & $0.82$ & $0.79$ \\
On-topic inclusion          & $0.88$ & $0.88$ & $0.85$ & $0.80$ & $0.80$ & $0.77$ \\
Topic ($4$-class)           & $0.93$ & $0.92^{m}$ & -- & $0.990$ & $0.993^{m}$ & -- \\
\bottomrule
\end{tabular}}
\end{table}

\section{Additional robustness}\label{app:robust}
\paragraph{Discovery and confirmatory windows.}
Table~\ref{tab:dc} reports the account-level reach contrast separately in the pre-specified discovery (3--15~Apr) and confirmatory (16--28~Apr) windows. The contrast is null and stable across both, with none of the strengthening under confirmation that a real effect would produce.

\begin{table}[h]
\centering
\caption{Discovery (3--15~Apr) $\rightarrow$ confirmatory (16--28~Apr) account-level reach
contrast. Null and stable across windows.}
\label{tab:dc}
\begin{tabular}{llccc}
\toprule
Topic & Window & $n_a/n_b$ & $\delta$ & $p$ \\
\midrule
TRUMP   & discovery    & 17/13 & $+0.22$ & $0.32$ \\
TRUMP   & confirmatory & 16/13 & $-0.04$ & $0.88$ \\
ISR/PAL & discovery    & 12/12 & $+0.17$ & $0.51$ \\
ISR/PAL & confirmatory & 13/12 & $+0.09$ & $0.72$ \\
\bottomrule
\end{tabular}
\end{table}

\paragraph{Targeted confound checks.}
Two targeted checks close the remaining confound channels. For \emph{age}, where the two sides enter the panel at very different points in a video's life, we recompute the account-level reach contrast on only the videos first observed within $24$h of posting (and within $48$h). The contrast stays null on every topic (Trump $\delta=-0.27$, $p=0.29$; Israel/Palestine $\delta=0.13$, $p=0.71$; ICE $\delta=0.20$, $p=0.79$ at the $24$h cut), so the age imbalance is not hiding a reach gap, though these subsets are small. For \emph{language}, ICE and Trump content is almost entirely English, so the Arabic confound is specific to Israel/Palestine; stratifying the other two topics by language leaves their reach nulls unchanged. Finally, because the account median is the primary estimand we report its support. The median account-topic unit rests on $5$ videos (range $1$--$164$), thin for ICE (mostly one to three videos per account, another reason we do not infer from it) and adequate for the informative topics.

\begin{figure}[t]
\centering
\includegraphics[width=0.96\columnwidth]{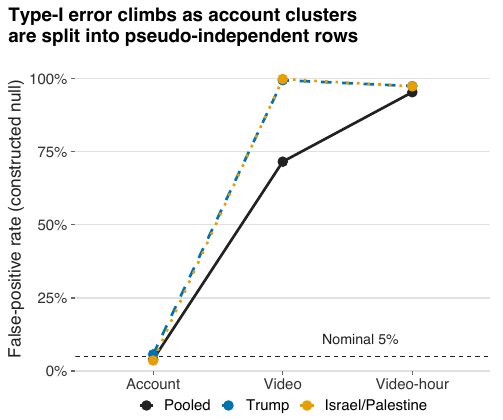}
\caption{False-positive rate as a function of the analysis unit, under a constructed
null with no true stance effect. Error is calibrated only at the account level and
climbs toward certainty as the unit is inflated.}
\label{fig:fprunit}
\end{figure}

\paragraph{Per-topic constructed null.}
Figure~\ref{fig:fprunit} (Appendix~\ref{app:figures}) shows the constructed-null false-positive rate at each unit of analysis, pooled and per topic. The naive pooled test rejects almost always; the account-level test is calibrated near the nominal $5\%$.

\paragraph{Multiple-testing sensitivity.}
Under the pre-registered Benjamini--Hochberg correction within outcome families, the engagement effect survives ($q=0.025$ for Trump, $q=0.021$ for Israel/Palestine). Under a correction across the six informative-topic cells it remains below $0.05$ ($q=0.049$ and $0.043$); under the strictest global correction across all nine grid cells it is borderline ($q=0.073$ and $0.064$). No reach cell is significant under any of the three families.

\section{Adjusted post-level model}\label{app:postlevel}
Table~\ref{tab:post_level_adjusted_model} details the per-video model of $\log(\text{peak plays})$ on curated side with $\log(\text{followers})$ and $\log(\text{video age})$ as controls, standard errors clustered by account (Section~\ref{sec:why}). Alongside the cluster-robust Wald $p$ we report a wild-cluster bootstrap $p$ (Rademacher weights, null imposed, $B=9{,}999$, seed fixed), the small-cluster-aware sensitivity appropriate at a few dozen clusters. No side coefficient is significant under either test, and the nominally borderline Trump cell ($p=0.07$) is clearly null under the bootstrap ($p=0.22$), consistent with the known anti-conservatism of cluster-robust Wald tests at these cluster counts.

\begin{table}[h]
\centering
\caption{Adjusted post-level model of $\log(\text{peak plays})$ on curated side, with
$\log$ followers and $\log$ video age as controls and account-clustered errors.
Coefficients are the log peak-play difference, establishment minus oppositional side.
\emph{Wild $p$} is a wild-cluster bootstrap (Rademacher, null imposed, $B=9{,}999$).
Cluster-robust SEs are anti-conservative at these cluster counts (especially ICE, $16$
clusters); no coefficient is significant under either test.}
\label{tab:post_level_adjusted_model}
\resizebox{\columnwidth}{!}{%
\begin{tabular}{llrrrrrrr}
\toprule
Topic & Side coding & Coef & Cluster SE & 95\% CI & $p$ & Wild $p$ & Videos & Clusters \\
\midrule
ICE & \textsc{pro}$-$\textsc{against} & 0.34 & 0.86 & $[\text{-}1.49, 2.16]$ & 0.70 & 0.69 & 28 & 16 \\
Trump & \textsc{pro}$-$\textsc{against} & 0.84 & 0.47 & $[\text{-}0.10, 1.79]$ & 0.07 & 0.22 & 355 & 37 \\
Israel/Palestine & \textsc{pro-isr}$-$\textsc{pro-pal} & 0.45 & 0.61 & $[\text{-}0.79, 1.69]$ & 0.46 & 0.56 & 371 & 32 \\
\bottomrule
\end{tabular}%
}
\end{table}

\section{Extended discussion}\label{app:extdisc}
None of these steps is novel on its own; each is standard in the literatures on clustered data and researcher degrees of freedom~\cite{lazic2018whatisn, simonsohn2020specification}. What is specific to visibility auditing is the order of operations. The unit must be fixed first, because it is the choice with the largest and least visible effect on the result, and the design effect must be reported alongside the $p$-value, because the two together are what distinguish an informative sample from a repeated one. A reader given only $N$ and $p$ cannot distinguish the dense audit we analyze from a thin one inflated by pooling, which is precisely the confusion on which the shadow-ban discourse depends.

\textbf{What would settle the question.} Our null is observational, and we read it as the absence of evidence at a level the design can resolve, not as proof of platform neutrality. The cleanest test is the one the 2020 Facebook and Instagram studies ran~\cite{guess2023feed,nyhan2023likeminded,gonzalezbailon2023asymmetric}, randomizing a ranking intervention for consenting users and measuring the downstream change. Where the amplification audit of \citet{huszar2022amplification} found a real but bounded partisan effect on Twitter using a randomized holdout, our account-level audit finds none for reach on TikTok, and the contrast is instructive precisely because both estimate effects at the account level. An external observational audit, however dense, should be read as a calibrated bound on the size of any effect, not as a verdict on intent.

\textbf{Toward auditable platforms.} The asymmetry we document, that a significant-looking result is easy to produce and a defensible one is costly, is ultimately an argument for structured data access. Independent verification of visibility claims needs sampling frames that fix account size, language, and region, and ideally platform-side randomization or logged exposure. Until such access is routine, the burden falls on analysts to report design effects, effective sample sizes, and full specification grids, so a reader can see whether a reported gap survives the choices that produced it.

\section{Reproducibility}\label{app:repro}
Every number in this paper is derived from a single statistics file produced by our analysis pipeline, and the figures are generated from that file and a small set of tidy per-account tables. The analysis grid (3 topics $\times$ 3 outcomes), unit, estimators, correction family, and discovery/confirmatory split were fixed in a pre-registration frozen before any inferential test; we report every cell with Benjamini--Hochberg correction and log all deviations against that plan. We will release the pre-registration, the de-identified analysis code, the figure-generation scripts, and the derived statistics (with an anonymized pre-registration reference during review) so that the full grid, the constructed-null calibration, and the specification curve can be reproduced and re-examined.

\section{Additional figures}\label{app:figures}

\begin{figure}[t]
\centering
\includegraphics[width=0.95\columnwidth]{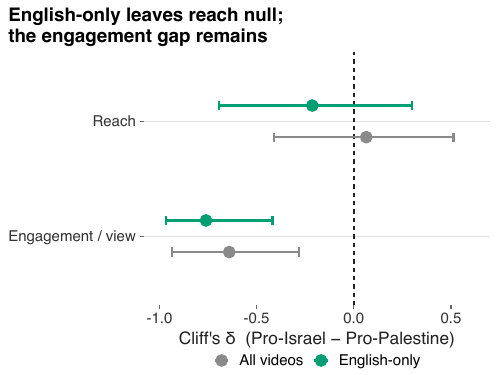}
\caption{Israel/Palestine account-level comparison, all videos vs.\ English-only.
The reach contrast is null in both; the engagement contrast (pro-Palestine higher)
persists.}
\label{fig:lang}
\end{figure}

\end{document}